# Ion Current as a Precise Measure of the Loading Rate of a Magneto-Optical Trap

W. Jiang,[1,*] K. Bailey,[1] Z.-T. Lu,[1,2] P. Mueller,[1] T. P. O'Connor,[1] R. Purtschert[3]

[1]Physics Division, Argonne National Laboratory, 9700 S Cass Ave., Lemont, IL, 60517, USA
[2]Department of Physics and Enrico Fermi Institute, University of Chicago, Chicago, USA
[3]Climate and Environmental Physics, University of Bern, CH-3012 Bern, Switzerland
*Corresponding author: wjiang@phy.anl.gov

(10/15/2013)

We have demonstrated that the ion current resulting from collisions between metastable krypton atoms in a magneto-optical trap can be used to precisely measure the trap loading rate. We measured both the ion current of the abundant isotope $^{83}$Kr (isotopic abundance = 11%) and the single-atom counting rate of the rare isotope $^{85}$Kr (isotopic abundance ~ $1\times10^{-11}$), and found the two quantities to be proportional at a precision level of 0.9%. This work results in a significant improvement in using the magneto-optical trap as an analytical tool for noble-gas isotope ratio measurements, and will benefit both atomic physics studies and applications in the earth sciences.

The magneto-optical trap (MOT) is widely used to produce cold (sub-mK) atoms for both scientific exploration and technical applications [1]. While the basic principle of the MOT is well established, the behavior of a real MOT can be difficult to characterize with a model from first principles [2,3], mainly because the atoms in a MOT are not a system of independent particles. Instead, they interact with each other by both collisions and reabsorption of fluorescence photons. These interaction mechanisms can be significant, even dominant, in a MOT with a readily achievable atomic density. In order to characterize the behavior of atoms in a MOT, various techniques have been developed to determine the key parameters, such as the number of atoms in a trap [4], the density [4,5], the temperature [6] and the loading rate [7,8], defined as the number of atoms being captured in a MOT per unit time.

In this paper, we report on a method of precisely determining the loading rate of a MOT of metastable noble-gas atoms. At a density greater than $10^{10}$ atoms cm$^{-3}$, the rate of atoms being lost from a metastable noble-gas atom trap is typically dominated by collision-induced ionization processes. At equilibrium, the ion current can be monitored to measure the trap loss rate and, equally, the loading rate. Instead of being a source of complication, the collisions between trapped atoms are found to be useful here in precisely determining a key parameter of the trap.

The demonstration of this method has been carried out with krypton atoms in the metastable state 5s[3/2]$_2$. A specific motivation for choosing Kr in this work is to use MOTs for isotope ratio measurements. The MOT is sensitive: a single trapped atom can readily be observed by its fluorescence. The MOT is also selective: once the laser frequency is chosen, only atoms of the desired element and isotope are trapped. Taking advantage of these unique properties of MOT, a method called Atom Trap Trace Analysis [9] has been used to measure extremely low isotope ratios, including the $^{85}$Kr/$^{83}$Kr ratios in the range of $10^{-12}$ [7,10] and $^{39}$Ar/$^{40}$Ar ratios in the range of $10^{-16}$ [11]. Each isotope-ratio measurement consists of two steps: first, the loading rate of the rare isotope (eg. $^{85}$Kr), on the order of 1 s$^{-1}$, is measured by counting single atoms in the trap; second, the loading rate of the abundant isotope (eg. $^{83}$Kr), at $10^{11}$ s$^{-1}$ or even higher, is measured. To obtain higher precision in the second step is the motivation of this work.

In two earlier methods of measuring the loading rates of the abundant isotope $^{83}$Kr [7, 8], the fluorescence of the $^{83}$Kr atoms in the trap is monitored either upon [7] or immediately following [8] the illumination with a resonant quench laser beam, resulting in the precision of each measurement to reach 5% and 2%, respectively. Both methods need a frequency-locked laser to excite the quench transition at a wavelength different from the trapping lasers in order to actively depopulate the metastable krypton atoms back to the ground state. Moreover, the alignment of the quench laser beam to the trap is critical. As a result, frequent calibration of the isotope ratio measurement using a standard gas sample is required. In comparison, the ion current method developed in this work avoids these complications. As a result, the measurements are significantly simpler and more reliable. Furthermore, the precision of each measurement is improved to 0.6%.

The equation governing the nonlinear dynamics of the number of atoms in a trap is given by:

$$\frac{dN(t)}{dt} = L - \gamma N(t) - \beta \int_V n(r,t)^2 dV \qquad (1)$$

Here N(t) is the number of trapped atoms and n(**r**,t) is the density distribution of the atomic cloud. On the right-hand side of the equation are three rates: the constant loading rate L; the background loss rate γ due to collisions between a trapped atom and the residual room-temperature gases in an imperfect vacuum and the ionization loss rate β due to cold collisions between two trapped atoms, which depends on the density profile of the atoms. The major ionization loss mechanisms are [12,13] (here Kr* denotes an atom in the metastable state):

Penning ionization, the dominant process
$$Kr^* + Kr^* \rightarrow Kr + Kr^+ + e;$$
associative ionization
$$Kr^* + Kr^* \rightarrow Kr_2^+ + e.$$

In principle, the loading rate (L) can be derived from the number of atoms at equilibrium measured by detecting fluorescence, and the trap lifetime measured by studying the decay of the atoms in the MOT after turning off loading. However, precision measurements using this approach is difficult because the lifetime changes as the density profile evolves. Here we investigate the possibility of measuring the loading rate of the MOT by collecting both the $Kr^+$ and $Kr_2^+$ ions generated from the cold collisions between trapped atoms. Under typical conditions in our experiments, the number of trapped atoms is on the order of $10^9$, the diameter of the atomic cloud is about 0.3 cm and the density can reach $10^{11}$ cm$^{-3}$. With a vacuum of $10^{-8}$ Torr, the loss rate due to collisions with background gas is on the order of 1 s$^{-1}$. Meanwhile, the ionization loss rate is on the order of $10^2$ s$^{-1}$, and exceeds the background loss rate by two orders of magnitude. Therefore, in steady state, the loading rate is equal to the ionization loss rate to a good approximation, and the ion current of $Kr^+$ and $Kr_2^+$ (The third term on the right side of Eq. (1)) can be a precise measure of the loading rate of the MOT.

In our MOT setup, the metastable Kr atoms are produced in a RF-driven gas discharge. Laser trapping and cooling are implemented by resonantly exciting the 5s[3/2]$_2$ – 5p[5/2]$_3$ transitions. The trapped atom cloud is surrounded by four C-shaped electrodes (Fig. 1), which are designed "to be out of the way", meaning not blocking any of the trapping laser beams. The electrodes are positively biased to a few kilovolts to generate an electric field of around 120 V/cm near the center of the MOT that guides the $Kr^+$ and $Kr_2^+$ ions toward a Faraday cup at the electrical ground. The neutral metastable Kr trap appears not to be affected by the electric field. Note that the Penning ionization and associative ionization processes are indistinguishable in our experiment because only the current is measured. The current, in the range of 0.1 - 2 nA, is processed through a low-noise current amplifier and recorded.

Figure 2 illustrates the relationship between the measured ion current from the MOT and the number of trapped atoms as measured by the fluorescence method. Given that the intensity of each trapping laser beam is about 50 times the saturation intensity, it is reasonable to approximate the number of atoms with fluorescence on this log-log plot even at the atom density of $10^{11}$ cm$^{-3}$. In the low-density limit, the atoms in the trap act as independent particles. The density distribution n(r) is then expected to increase proportionally with the total number of trapped atoms N. Consequently, the ion current, caused by the ionization loss term in Eq. (1), is expected to be proportional to $N^2$. However, the data plotted in Fig. 2 clearly deviates from the simple $N^2$ dependence illustrated by the dashed line. At these densities, the reabsorption of fluorescence photons among the trapped atoms results in an effective repulsive force that pushes the atom cloud apart [14], and modifies the density profile. Indeed, the volume of the trap grows as the number of atoms increases while the central density remains saturated. In a previous work [2], as many as four different MOT operating regimes were used to model the dependence.

We use an empirical model to describe the evolution of the density profile. The density distribution of the atoms inside the MOT is assumed to be spherically symmetric, and follow the Fermi function (see inset of Fig. 2),

$$\rho(r) = \frac{n_0}{1 - a_1 \exp\left(\frac{r-R}{a_2}\right)} \qquad (2)$$

where $R$ is the radius of the atom cloud; $a_1$ and $a_2$ are fitting parameters, with $a_2$ often described as the "skin depth"; $n_0$ is the density at the center. Given a set of fit parameters, the number of atoms in the trap (N) is calculated by integrating $\rho(r)$ over the trap volume; the ion current is calculated by integrating $\rho(r)^2$ over the trap volume according to the ionization loss term in Eq. (1). Fermi function is chosen here because it conveniently takes into account the saturation of the central density as the trap grows. The solid line in Fig. 2 shows that there is a qualitative agreement between the data and the Fermi function model.

Even though the relationship between the current and the number of atoms is nonlinear and complex, we demonstrate below that, in stark contrast, the current and the trap loading rate are strictly proportional. For this demonstration, the ion current of the abundant isotope $^{83}$Kr (isotopic abundance = 11%) is compared to the loading rate of the rare radioisotope $^{85}$Kr (isotopic abundance = 1×10$^{-11}$) at identical experimental conditions. While the laser frequency is switched back and forth between the resonances of the two isotopes every few minutes, laser power, alignment, magnetic fields, the gas flow rate, RF power, etc., are all kept unchanged. Due to its extremely low isotopic abundance, $^{85}$Kr atoms are present in the MOT in numbers of 0, 1, or 2, which can be clearly resolved by the discrete levels of fluorescence (Fig. 3). The loading rate of $^{85}$Kr is determined by the atom counting rate, and is expected to be proportional to the loading rate of $^{83}$Kr with the proportional coefficient given by the isotope-abundance ratio of a given sample.

Two types of comparison are performed to test the new method. First, the isotope ratio of a modern atmospheric Kr sample was measured repeatedly over a period of two weeks. Various experimental parameters, such as the intensities and alignment of both the trapping and slowing laser beams, the RF power used to drive the discharge, etc., were deliberately adjusted between measurements in order to change the ion current by more than an order of magnitude in the range of 0.1 – 2 nA. Figure 4 shows the correlation between the ion current of $^{83}$Kr and the atom counting rate of $^{85}$Kr. Data analysis indicates that it is a proportional relationship with $\chi^2/\nu = 1.2$, and that the uncertainty of the proportional coefficient is 0.9%. For each measurement, the statistical uncertainty for $^{85}$Kr counts is typically about 3%, corresponding to approximately 1000 counts. The uncertainty for the current of $^{83}$Kr is at 0.6%, thanks to the excellent signal-to-noise ratio. The results support the fact that the ion current is a robust measure that is strictly proportional to the trap loading rate. As the ion current drops somewhat below 0.1 nA, the ionization loss rate is still expected to dominate, thus the proportional relationship is likely to continue perhaps all the way down to 0.01 nA. However, relationship over this lower current range has not been experimentally verified due to the poor counting rates of $^{85}$Kr.

The second comparison aimed to check the linearity of the isotope-ratio measurements. A group of five krypton samples with different $^{85}$Kr/$^{83}$Kr ratios were produced at the University of Bern by mixing a modern atmospheric sample ($^{85}$Kr/$^{83}$Kr = 2×10$^{-10}$) and an old atmospheric sample ($^{85}$Kr/$^{83}$Kr < 10$^{-12}$). Also at the University of Bern, the resulting $^{85}$Kr/$^{83}$Kr ratios, in the range of 10$^{-12}$ - 10$^{-10}$, were determined both by the known volume-mixing ratios and by Low Level Decay Counting (LLC) of $^{85}$Kr (half-life = 10.8 yr) [7,15]. We then compared the Bern results with the isotope ratio measurements conducted at Argonne using the ion current as a measure of the trap loading rate of $^{83}$Kr. The two independent measurements conducted in separate laboratories agree at the 3% level ($\chi^2/\nu = 0.8$) (Fig. 5).

With these demonstrations, the ion current method for measuring the loading rate of a MOT of metastable noble-gas atoms is validated. This work enables an advance in both operational simplicity and reliability of Atom Trap Trace Analysis, and improves the accuracy of isotope-ratio measurements. Indeed, the new method has been implemented at Argonne, and is used routinely to analyze environmental samples for applications of radioisotope dating in the earth sciences [16].

We thank S.-M. Hu and H. Esbensen for stimulating discussions. This work is supported by the Department of Energy, Office of Nuclear Physics, under contract DE-AC02-06CH11357.

# Figures

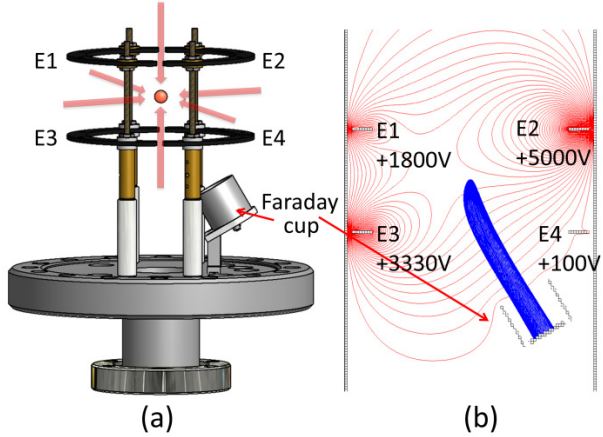

Fig. 1. Four biased electrodes (E1 to E4) guide the $Kr^+$ and $Kr_2^+$ ions, produced by collisions between trapped atoms (red dot), toward the Faraday cup. (a) Schematics of the setup. (b) SIMION® calculations of the electric potential (red lines) and simulations of the ion trajectories (blue lines). The thick patch of blue lines is formed by a range of trajectories due to the initial distribution of positions and momenta. The initial kinetic energy of the Kr ions is set to be 5 eV, which is close to the maximum kinetic energy the $Kr^+$ and $Kr_2^+$ ions can get in the ionization processes.

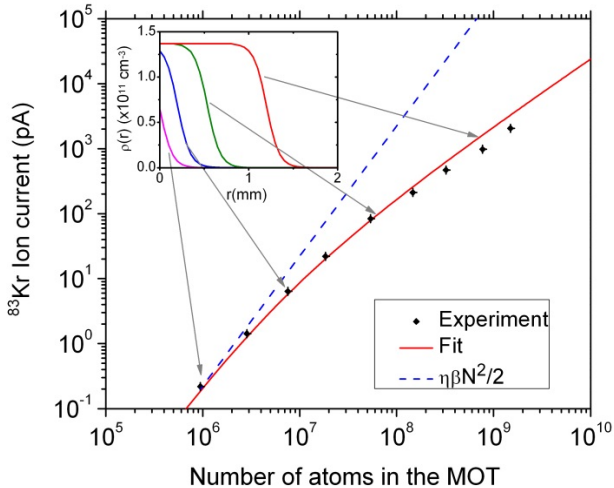

Fig. 2. $^{83}Kr$ ion current versus the number of trapped atoms. The dash line corresponds to the simple $N^2$ dependence. $\eta$ is the ion collection efficiency; $\beta$ is a coefficient defined in Eq. (1). The solid line shows the results of the Fermi function model. Inset: Density distributions of atoms in the trap. Various curves correspond to different total number of atoms in the trap.

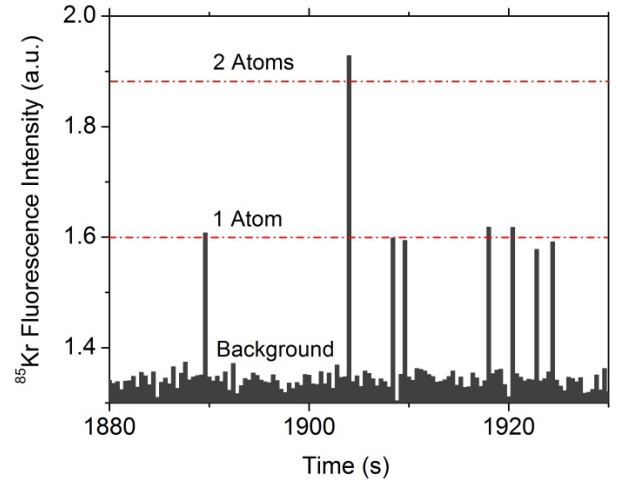

Fig. 3. Individual $^{85}Kr$ atoms can be counted by its discrete levels of fluorescence. The background is photon counts from light scattered off surrounding walls and optics.

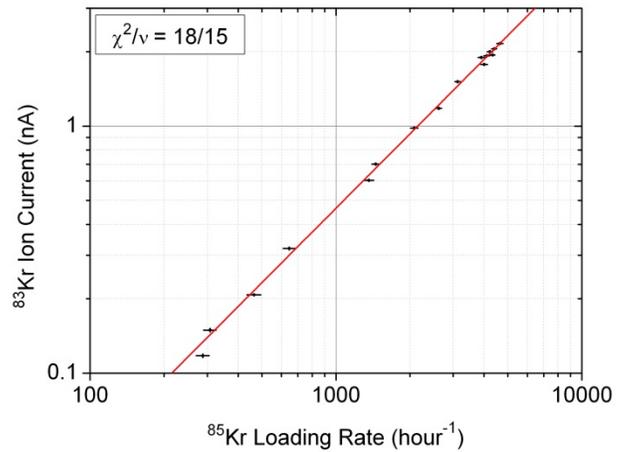

Fig. 4. $^{83}Kr$ ion current versus $^{85}Kr$ loading rate. Both are varied deliberately by more than an order of magnitude. The proportional relationship proves that the ion current is a robust measure of the trap loading rate.

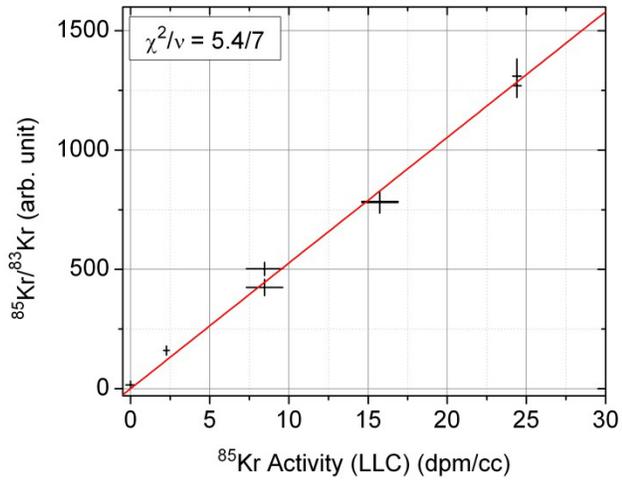

Fig. 5. $^{85}Kr/^{83}Kr$ versus $^{85}Kr$ activities (Low-Level Decay Counting). The standard units for $^{85}Kr$ activity is dpm/cc: decays per minute per cubic-centimeters of Kr gas at the standard temperature and pressure (100 dpm/cc corresponds to an isotopic abundance of $3\times10^{-11}$). The agreement between the results measured by two completely different methods provides further validation to the ion-current method.